\documentclass[aps,prb,reprint,showpacs,superscriptaddress,groupedaddress]{revtex4-1}
\usepackage{graphicx}
\usepackage{amsmath}
\usepackage{amssymb}
\usepackage{dcolumn}
\usepackage{latexsym}
\usepackage{rotating}
\usepackage[usenames,dvipsnames]{xcolor}
\usepackage{float}
\usepackage{epsfig}
\usepackage{psfrag}
\usepackage{natbib}
\usepackage{bm}
\usepackage{eucal}
\usepackage{braket}
\usepackage{enumerate}
\usepackage{longtable}
\usepackage{subfigure}
\usepackage{bm}
\usepackage{hyperref}
\usepackage{amsfonts}
\usepackage{dcolumn}
\usepackage{bm}
\usepackage{subfigure}

\newcommand{\be}{\begin{equation}}
\newcommand{\ee}{\end{equation}}
\newcommand{\bn}{\begin{eqnarray}}
\newcommand{\en}{\end{eqnarray}}

\usepackage{color} 

\usepackage{hyperref}
\begin{document}

\author{S. Koley$^{1}$}\email{sudiptakoley20@gmail.com}

\title{Theoretical study on spintronic and optical property prediction of doped magnetic Borophene}
\affiliation{$^{1}$ Department of Physics, Amity Institute of Applied Sciences, Amity University Kolkata, West Bengal, 700135, India}
\begin{abstract}
Two dimensional materials are attracting new research for optoelectronics and 
spintronics due to their unique physical properties. A wide range of emerging 
	spintronic devices are achieved from parent and doped two dimensional materials. 
First-principles electronic structure calculations of a two- 
dimensional  monolayer of borophene in its P6/mmm form is explored in this study. The 
electronic, magnetic, and optical properties of doped borophene are analyzed
	for 
doping with lithium, sodium, and magnesium. Density functional theory 
calculations advocate their good 
	dynamical and thermal stability. Spin-polarized electronic properties of
these materials are observed to be useful for predicting new spintronic 
	materials. Additionally optical analysis exhibits the absorption peaks 
	in monolayers along the in-plane direction. These properties of doped 
	magnetic borophene suggest the material to be a suitable candidate for 
	designing optoelectronic devices. The most competent spintronic material
	 among three different doped borophenes is lithium doping that can imply a promising avenue for the fast-growing electronics industry.
\end{abstract}
\maketitle

\section{Introduction}
\noindent Two-dimensional (2D) materials have been celebrated since the first synthesis of 
graphene as it displays exceptional conductivity properties which are lacking in
its bulk crystal form\cite{novo, avouris}. The 2D material demonstrates
a wide range of unique optical, electronic,
 and thermal properties, hence has enormous applications in recent technology 
\cite{mak,wang,shin}. 
A considerable amount of research continues to explore new 2D materials: silicene, 
phosphorene, borophene, \cite{randivir,rubab,molle,kou,chen1,zhang1}chalcogenides\cite{sk1,sk2,sk3,sk4} and layered binary compounds 
like GeP, GeAs, SiP\cite{beck,young}.
The common structural property among these materials is extreme thinness which 
dramatically changes the physical properties as compared to the corresponding bulk 
materials.

\noindent Borophene, a single layer of boron atom, has been synthesized at laboratory in 
different phases like Pmmn, $\beta_{12}$, $\xi_3$\cite{1903.11304}. Strength, 
flexibility, and metallic attributes of this material have initiated ample research on their mechanical 
properties, thermal conductivity, electronic structure stability, superconductivity,  
optical properties, and atomic adsorption \cite{1903.11304}. Depending on the 
structural 
orientation and doping, borophene turns out to be a good conductor of heat and electricity. Additionally phonon-mediated superconductivity is observed at a 
$T_C$ of 25.3K. A more remarkable discovery was 
a two-dimensional phase of boron with space group Pmmn showing tilted Dirac cone in its 
electronic band structure\cite{PBL_1608.05018}. Borophene has low atomic weight and excellent electrical performances; moreover it's source is less expensive. 
 Overall the structure is tunable which escalates its practical 
applications and research interest.

\noindent Boron has the electronic structure [He]2s$^2$2p$^1$ with three valence 
electrons and is located in between nonmetallic carbon and metallic beryllium in the periodic
table. The electronic configuration itself provides both metallicity and nonmetallicity 
owing to the radius of 2p orbit which is close to the 2s orbit radius. Hence, 
the rich electronic structure and bonding between the boron atoms make its 2D structure 
different from other 2D materials. However the magnetic phase diagram and doping 
dependence on borophene remains largely unexplored. Amidst different structures of 
borophene only the P6/mmm structure shows suitable spin polarization. With the goal of 
uncovering 2D magnetic materials for superior spintronic and optical 
applications, the investigated structure is doped with both electron and hole. Present study quantifies
the optical and spintronic properties of borophene for different doping. Spintronic application of
doped borophene then can be used as new zero gap spintronic materials.

\begin{figure}
\epsfig{file=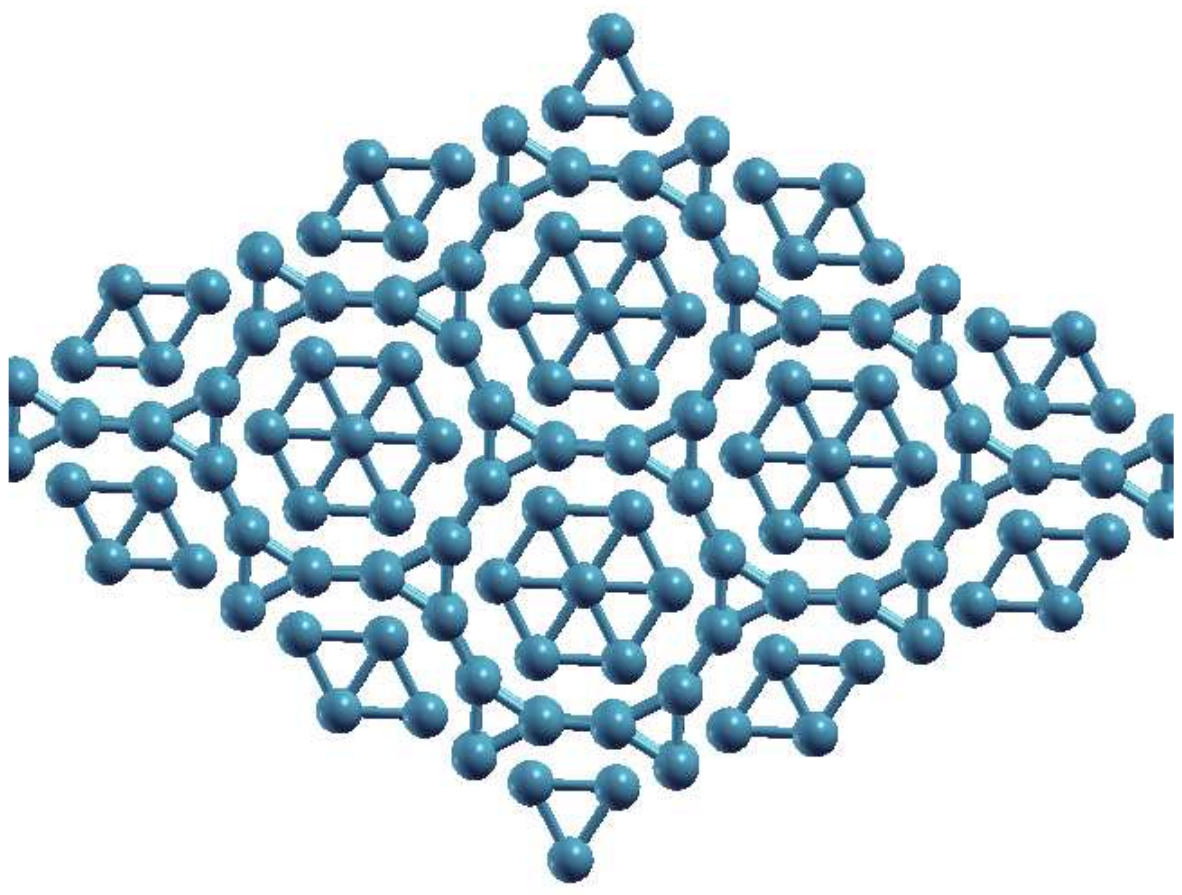,trim=1.8in 2.5in 1.5in 1.5in,clip=false, width=40mm}
\caption{(Color Online) Crystal structure of pristine borophene in P6/mmm phase with three inequivalent boron atoms.
}
\label{fig1}
\end{figure}

\begin{figure}
\epsfig{file=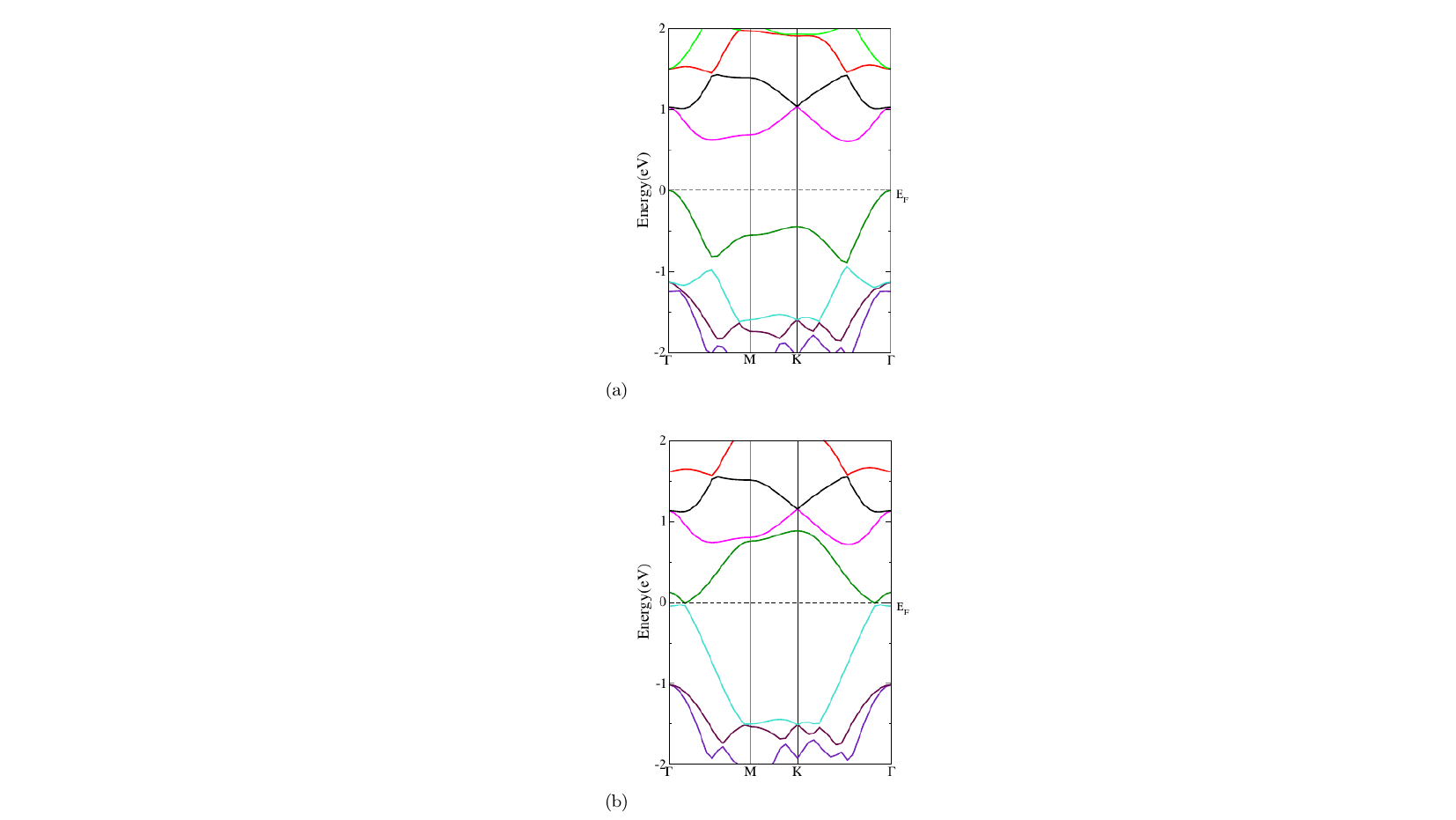,trim=4.5in 0.0in 2.5in 0.0in,clip=false, width=120mm}
\caption{(Color Online) Density functional theory band structure of parent borophene for (a) up-spin and (b) down-spin. Up-spin shows a comparably larger gap than down spin.}
\label{fig2}
\end{figure}
\begin{figure*}
\epsfig{file=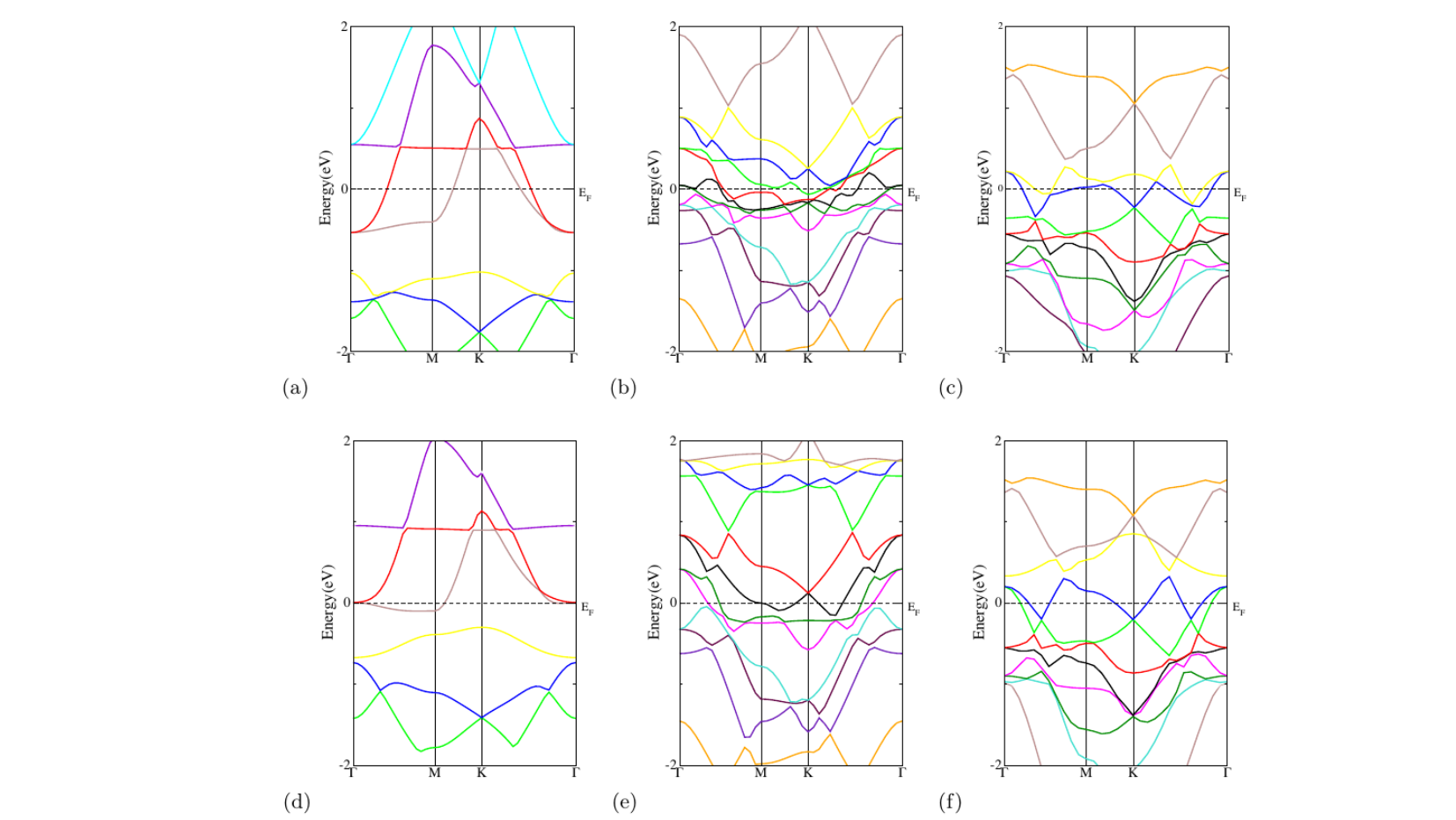,trim=1.0in 0.0in 0.0in 0.0in,clip=false, width=200mm}
\caption{(Color Online) Spin-polarized DFT band structures for Li((a) and (d)), Na((b) and (e)), and Mg((c) and (f)) doping. (a),(b) and (c) is for the up spin and others are for the down spin. 
The six plots are for doping concentrations of 20 $\%$.}
\label{fig3}
\end{figure*}

\section{Methods}
Electronic structure of borophene was estimated using density functional theory 
 calculations based on the Perdew-Burke-
Ernzerhof (PBE) generalized gradient approximation (GGA) for  
exchange-correlation functional, along with full potential linearized augmented 
plane wave(FP-LAPW) method as 
executed in WIEN2K\cite{w2k}. A kinetic energy cutoff of 20 eV was applied for the plane waves to get the electronic wave functions. 
The k-space integration was performed using a 
$10\times10\times1$ Monkhorst-Pack k-point mesh centered at $\Gamma$-point. 
Atomic 
coordinates and lattice constants were entirely relaxed until the forces 
were smaller than 0.001eV/$\AA$. 
The optical terms, such as imaginary and real
parts of dielectric 
tensor, and reflectivity were calculated within the WIEN2K code. The 
interstitial wave
functions were expanded in terms of plane waves 
with a cut-off parameter of R$_{MT}\times$ $K_{max}$=8.5, where R$_{MT}$ and 
K$_{max}$ denote the smallest
atomic sphere radius and the largest k vector in
the plane-wave expansion, respectively. The optical terms were evaluated using a 
dense k-point mesh of $15\times15\times1$ $\Gamma$-centered Monkhorst-Pack with 
fixing Lorentzian 
broadening of 0.05 eV. 

\section{Results}
In accordance with structural symmetry of P6/mmm borophene the unit cell consists of three 
types of inequivalent boron atoms. Three inequivalent atomic positions of boron 
are B1(0.0,0.0,0.5), B2(0.362,0.181,0.5), and B3 (0.576,0.153,0.434). The
lattice parameters are chosen to be a = b = 11.40 $\AA$, and c = 37.73 $\AA$. 
Figure.1 shows the crystal structure of parent 
borophene. The hexagonal unit cell is doped with other atoms for hole and 
electron doping. 
Doping site is chosen to be (0.0,0.0,0.5) point with 6/mmm symmetry for an 
energetically favourable scenario (in all the position stability of the atom 
is checked through scf calculated total energy, finally minimum energy structure is chosen for further calculation.)
A phonon spectra that shows real frequency dispersion was
achieved using the force-constant method and the
PHONOPY package\cite{phonopy}. This proves the
stability of the doped material (shown in supplementary data).
The different local environment of distinct boron atoms 
predicts that they can exhibit different chemical and physical properties.

The inequivalence of the boron atoms changes the contribution of 
each atom to the electronic properties of the material. Figure.2a and 2b demonstrate 
the electronic band diagram of parent borophene in a spin-polarized case (spin polarized density 
of states for parent material is attached in supplementary material). For 
up spin (Figure.2a) the band diagram shows a prominent indirect band-gap of 0.5 eV while for
a down spin (Figure.2b) the conduction band and valence band shows tilted Dirac 
 type 
 structure close to high symmetry $\Gamma$ point in reciprocal space. 
 Linear dispersion around Fermi energy of the down spin band diagram confirms 
 characteristic features of 
Dirac semimetals. Thus, the parent borophene in ferromagnetic
 structure is a direct narrow-gap semiconductor as the down spin conduction band
 and valence band are close to Fermi level.
An earlier calculation for magnetic borophene\cite{zhu} corroborates with the 
band structure of parent borophene. 

\begin{figure*}
\epsfig{file=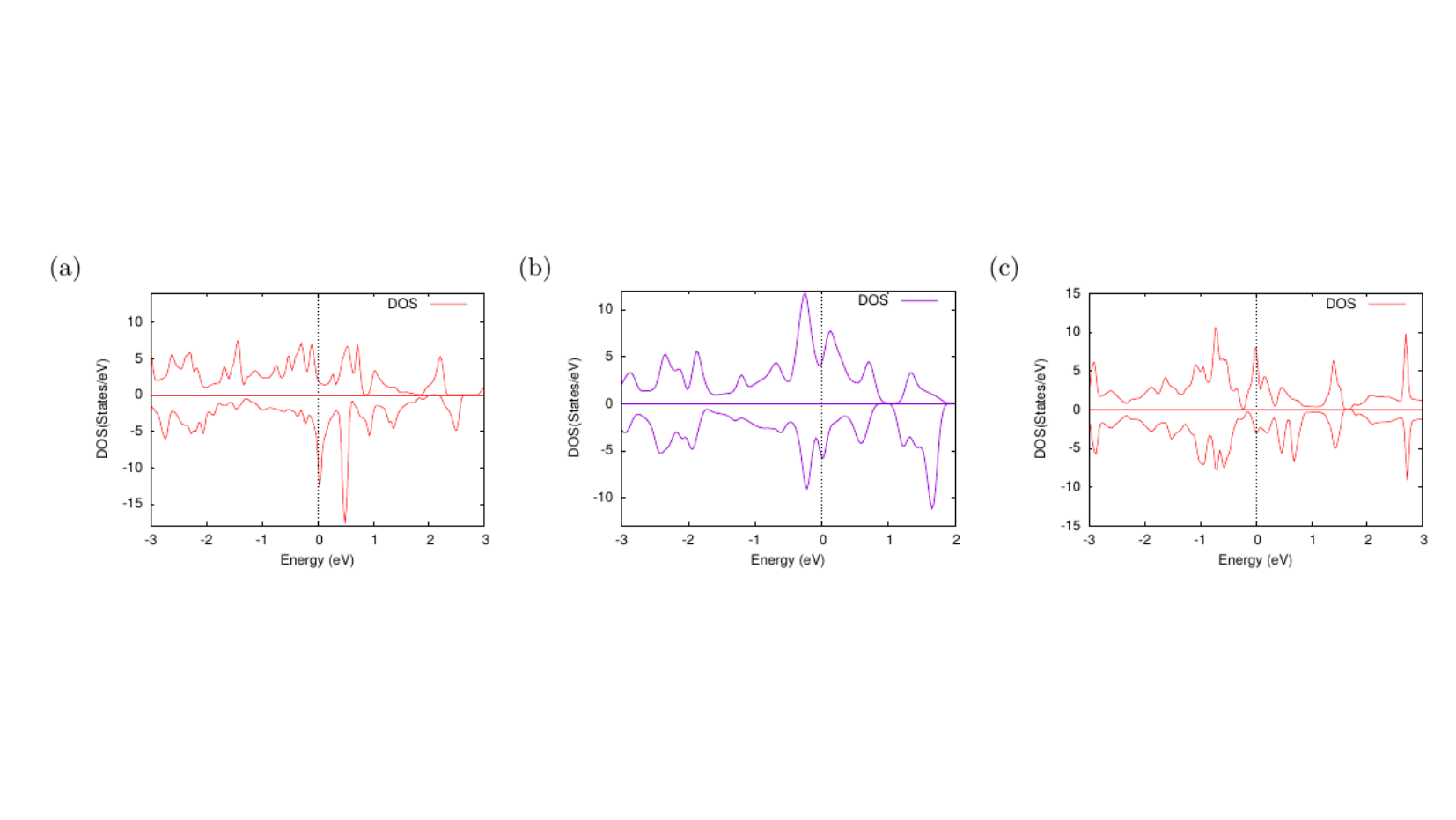,trim=0.0in 2.0in 0.0in 0.0in,clip=false, width=180mm}	
\caption{(Color Online) Spin-polarized DFT density of states for three dopants,
	(a) Li doping, (b)Na doping, (c) Mg doping.
The down spin density of states is multiplied by -1 for presentation.}
\label{fig4}
\end{figure*}
\begin{figure}
\epsfig{file=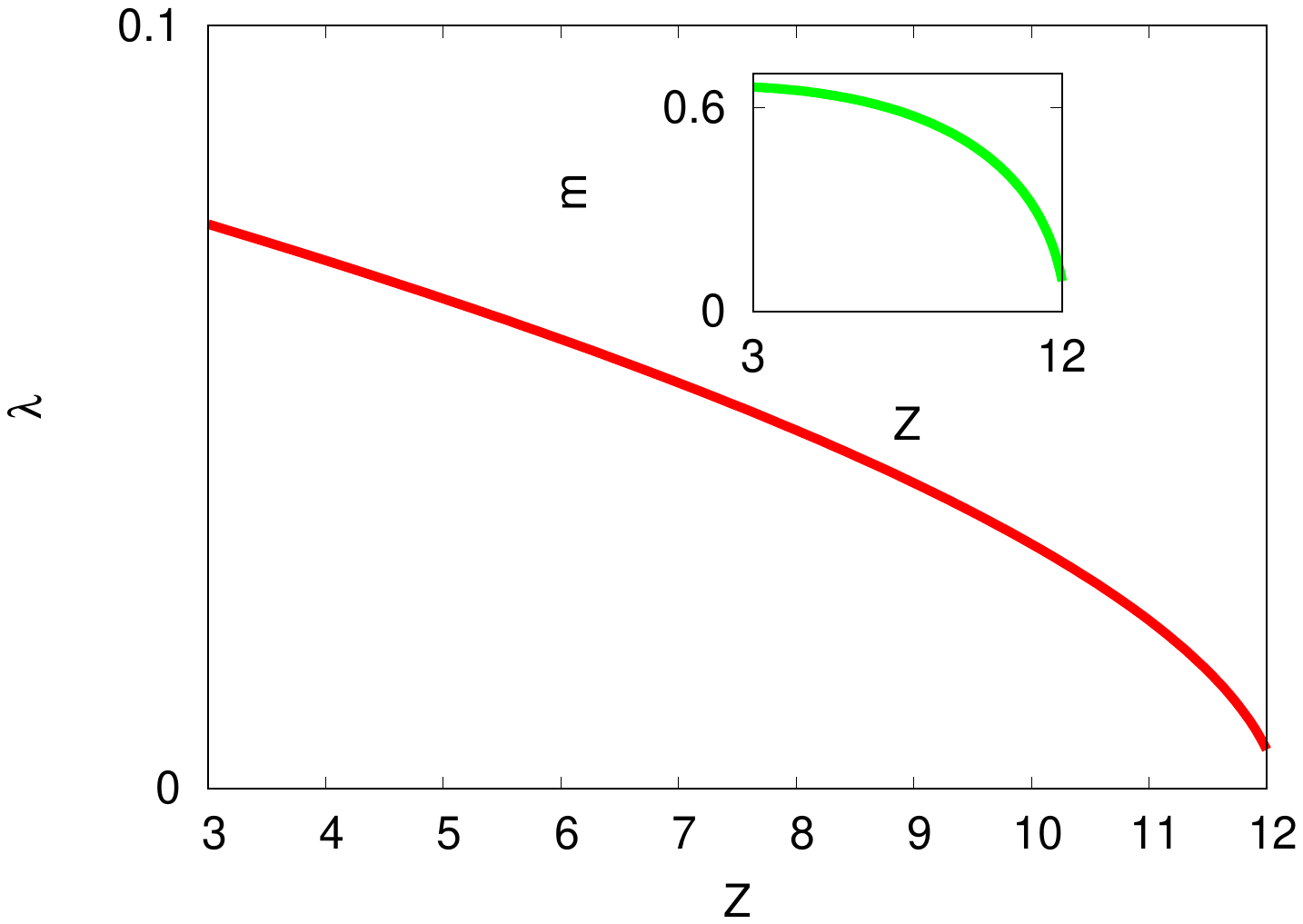,trim=1.0in 0.2in 0.0in 1.0in,clip=false, width=100mm}
\caption{(Color Online) Spintronic order parameter plot with atomic number. 
Inset shows the plot of magnetization (defined as the difference in number densities of up and down spin) with atomic number. 
It shows decrease in the order parameter and magnetization with increasing atomic number.}
\label{fig5}
\end{figure}

\begin{figure*}
\epsfig{file=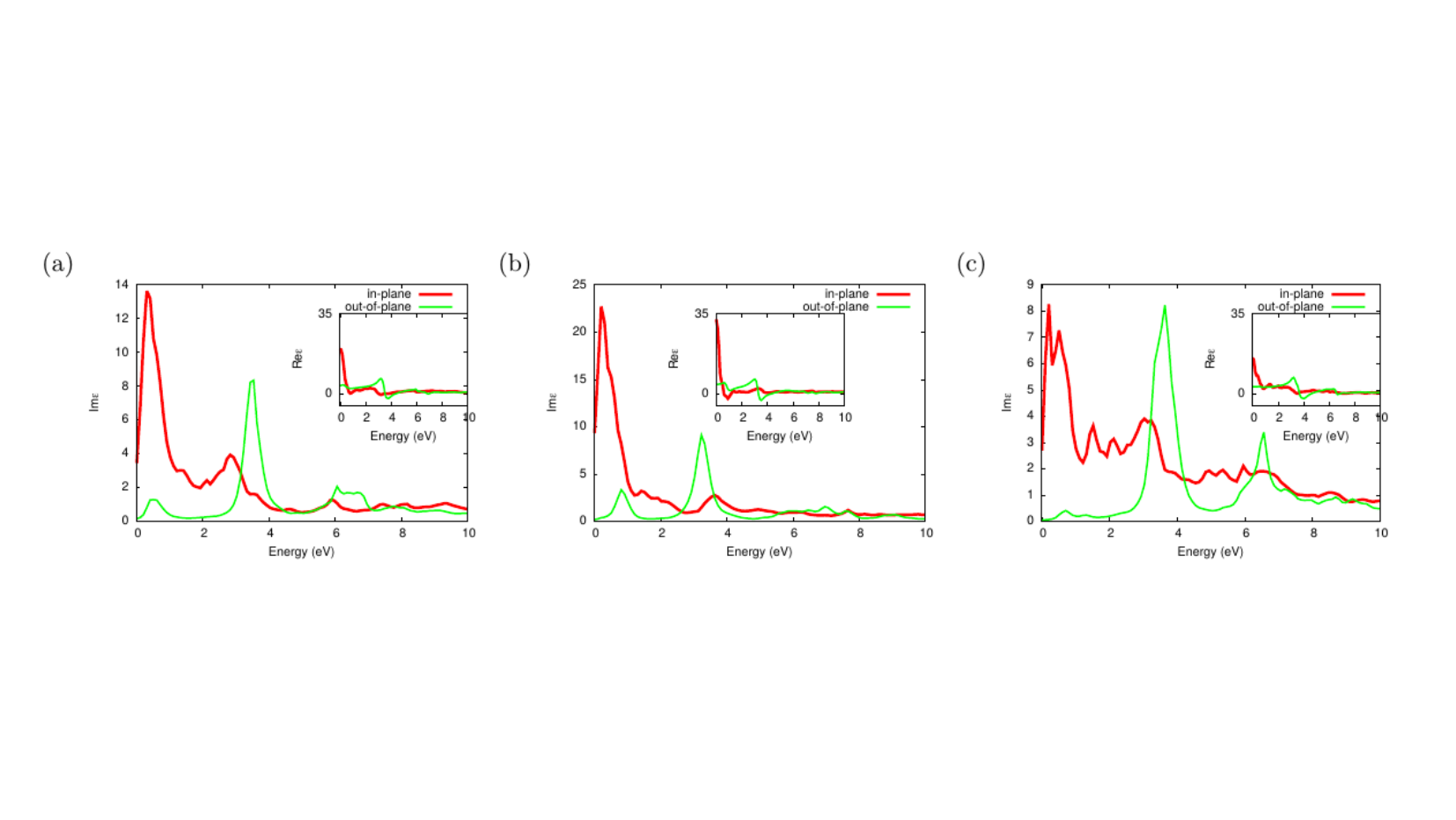,trim=0.9in 0.9in 0.9in 0.7in,clip=false, width=180mm}	
\caption{(Color Online) The imaginary part of the dielectric function 
obtained from DFT calculation for three dopants, (a) Li, (b)Na, and (c) Mg.
The in-plane and out-of-plane contribution is shown in all three cases.}
\label{fig6}
\end{figure*}

Spin-polarized band structure and density of states of three doped systems (i.e. Li, Na, and Mg 
) is plotted in Figure.3(a-f) and Figure.4(a-c) (partial density of states for three doping is 
included in supplementary section). 
The parent band structure establishes that the system can be tuned from semiconductor to half metal and semiconductor to metal with suitable doping. 
Band structures of doped borophene reveals that the doped systems show 
metallicity owing to the 
bands having finite weight at 
the Fermi level. The distinct behavior in electronic band structures can 
be originated from the electronic structure of the 
foreign elements and their corresponding positions in the periodic table. 
Compared to boron; lithium, and sodium have two fewer, and magnesium has one 
fewer valence electrons and they are adjacent to 
each other in the periodic table. 
As sufficient electrons are not present to form required covalent bonds surrounding the atoms, 
the studied new structures can be described as hole doped systems. A doping of hole type 
shifts the Fermi energy downward into the valence band and  Dirac
 nature can be observed above the Fermi level only. While there are 
 two structurally symmetric Dirac 
 cones present in the down spin band diagram of parent borophene, the Dirac 
 structure gets shifted in all the doped borophene.  Additionally the shifting 
 of the Dirac point depends on the type of dopant atoms. The lattice symmetry 
 breaking depends on the atomic radius difference between the dopant and parent atom. These findings suggest parent and doped borophene in this structure can be 
 a promising candidate for spintronic applications. 
\section{Spintronic properties}
An ideal spintronic material is selected by the number of spin-polarized carriers of a system. 
These properties were quantified by the number of spin-up and spin-down charge 
carriers in the doped borophenes. In this case, spintronic order 
parameter ($\lambda$) is defined as follows to measure the degree of 
spin polarization and compare them for finding most competent dopant, 
\begin{equation}
\lambda = \frac{n_{\uparrow} - n_{\downarrow}}{n_{\uparrow} + n_{\downarrow}}
\end{equation}
where $n_{\uparrow}$ and $n_{\downarrow}$ are the numbers of carriers respectively for up and down spins.
Large spin polarization demands $\lambda$ to be higher and thus make a suitable candidate 
for a spintronic system.

The main panel of figure.5 portrays $\lambda$ for three doped systems which
decreases with an increase in atomic number of dopants (Figure.5 main panel). 
This property of dopant implies elements having lower atomic number than boron can be employed to improve spintronic properties. The  
calculated $\lambda$ shows a smaller value that can be increased  
with external parameters such as pressure and biasing.
In this work, the energetically favorable structure is selected for these doped 
structures, however further research is required to understand whether other elements can yield similar energetically favorable 
scenario. Albeit the parent borophene in its crystal structure shows magnetism 
and is a satisfactory 
element for spintronic applications, the present study suggests doping will 
enhance the applicability. 

Doped borophene does not have any gap in its electronic band structure.
Similar results were obtained in earlier work on the intercalated 
phosphorene\cite{pho} and other zero gap materials\cite{X-L-wang, yue}. 
Besides, the doped borophene also has the 
difference between the number densities of the 
major and the minor spin states, which give rise to 
magnetic properties in the system. In the inset of Figure.5, the 
magnetization, $m = (n_{\uparrow} - n_{\downarrow})\mu_B$ is plotted as a 
function of atomic number. 
The trend of m shows a decrease in magnetization values with atomic number.

The concept of a gapless spintronic semiconductor\cite{X-L-wang} material is well established and finding new gapless spintronic materials is an emerging research area.
This study aims to get new spintronic material.
Doped borophene showing zero-gap state, are superior to  
pristine borophene since zero energy is required to 
excite electrons from the valence band to the conduction band in doped system.

\section{Optical properties}
Motivated by semiconducting behavior of parent material, the optical properties of doped borophene are studied to predict 
optoelectronic application of these materials. The imaginary part of 
dielectric function was estimated by applying a  
density functional theory method constructed
 through WIEN2K software. The assessed electronic structures of doped borophenes 
 discussed earlier in the results are the basis to interpret the 
 imaginary part of the 
 dielectric functions for three different systems. The optical
terms for polarization of light parallel to the plane are dominant because of
the strong depolarization effect in 2D planar geometry. The structures of three
doped borophenes
have geometrical symmetry
along the x- and y-axis. The imaginary part of the dielectric function 
can also be defined as optical absorption spectra
of these doped borophenes for the
in-plane and out-of-plane polarized directions which is illustrated in 
Figures.(6a-6c). The corresponding real parts 
of the dielectric functions are shown in the inset. The first peak of 
in-plane dielectric
spectra occurs around 0.5 eV for the Li, Na, and Mg doped borophenes.
Consequently, the first peaks of absorption spectra 
for all doped systems are in the near-infrared range of wavelength along the 
in-plane direction. These observation confirms that, the excitations for 
in-plane polarization are largely found within the
low-frequency region, whereas, the excitations for perpendicular polarization, 
are found within the higher frequency regime. The similar observation was 
reported earlier for parent borophene\cite{45chodhury}.

\section{Summary and Conclusions}
This study demonstrates density functional theory study of the magnetic 
and optical properties of Li, Na, and Mg-doped borophene in spin-polarized system. 
Borophene is a member of newly found 2D Dirac systems in its different 
structures and  has potential application because of its high capacity, outstanding electronic and 
ionic conductivity in metal ion batteries as an anode material in its different 
structures; however the research interest on its application as the spintronic 
materials is recent. 
Parent borophene in this p6/mmm structure shows half-metallic nature with up-spin 
showing semiconducting nature and down spin showing metallic behavior. Doping 
induces vanishing of gap but doped borophene remains spin-polarized. While 
the main characteristics are common in all doped crystals but the nature of spin
 polarization varies with different dopant. The detailed assessment into  
 spin-dependent properties shows magnetization, and reveals that 
 spintronic order parameter($\lambda$) decreases with increasing atomic number.
 In addition the imaginary part of the dielectric functions of
 all three doped materials are characterized by in-plane maxima around 0.5 eV 
 (corresponding to near-infrared region)
 and out-of-plane maxima around 3.0 eV (corresponding to visible region). 
 The lowest energy excitations demonstrate almost similar 
 characteristics in all the doped materials. 

In conclusion, Li, Na and Mg doping of borophene shows similar electronic and 
optical properties. The optical absorption threshold of all the materials 
shows promising applications because of photoemission in visible and near- 
infrared region. 
Spintronic properties reveal better performance for 
doping with lighter atoms which labels Li as the best among three. Thus this study suggests the potential 
for fine tuning the properties of borophene using doping.

\section{Conflict of Interest}
The author declares no conflict of interest.
\section{ACKNOWLEDGEMENT} 
SK acknowledges the Department of Science and Technology women scientist grant SR/WOS-A/PM-80/2016(G) for financial support.

\end{document}